# Tailoring of ferromagnetic $Pr_{0.85}Ca_{0.15}MnO_3$/ferroelectric $Ba_{0.6}Sr_{0.4}TiO_3$ superlattices for multiferroic properties

P. Murugavel, D. Saurel, W. Prellier[a], Ch. Simon, and B. Raveau

Laboratoire CRISMAT, CNRS UMR 6508, ENSICAEN, 6 Bd du Maréchal Juin,

F-14050 Caen Cedex, FRANCE.

Abstract

Superlattices composed of ferromagnetic $Pr_{0.85}Ca_{0.15}MnO_3$ and ferroelectric $Ba_{0.6}Sr_{0.4}TiO_3$ layers were fabricated on (100) $SrTiO_3$ substrates by a pulsed-laser deposition method. The capacitance and resistive parts of the samples were analyzed from the complex impedance measurements, performed on the samples using a special experimental set-up. The superlattice with larger ferroelectric thickness shows unique characteristics which are not present in the parent ferromagnetic thin film. The superlattice show both ferromagnetic and ferroelectric transitions which is an evidence for the coexistence of both the properties. The high magnetoresistance (40 % at 80K) shown by the superlattice can be attributed to the coupling between ferromagnetic and ferroelectric layers, i.e, to the magnetoelectric effect.

[a)]Electronic mail : prellier@ismra.fr

There has been growing interest in the study of magnetoelectric (ME) materials, a special class of mutiferroics in which two or three of ferroelectric, ferromagnetic, and ferroelastics properties coexist[1]. In the multiferroics with ferromagnetic and ferroelectric ordering simutaneously, one can expect the coupling between the magnetic and dielectric properties as well as their control by the application of magnetic and/or electric fields[2]. This property can be used as a basis for novel actuators and sensors with high sensitivity. However, there are very few natural multiferroic magnetoelectrics that are both ferromagnetic and ferroelectric in the same phase[3]. This is because, off-center distortion responsible for polar behavior is usually incompatible with the partially filled d-level which are prerequisite for a magnetic ground state[4]. A strong ME effect, however, could be realized in the composite consisting of magnetorestrictive and piezoelectric effects.[5] Most studies in the past focused exclusively on ferrite-(Pb,Zr)$TiO_3$/$BaTiO_3$ composites[5-9]. The mixed oxides yielded ME coefficients much smaller than calculated values due to leakage currents through low-resistivity ferrites and microcracks that resulted from mismatch of structural parameters and thermal properties. There have also been reports of magnetoelectric effects in layered composites such as (PZT)-$Tb_{0.3}Dy_{0.7}Fe_{1.92}$ (Terfenol-D), polyvinylidenefluoride-Terfenol-D and $Pb(Mg_{1/3}Nb_{2/3})O_3$-$PbTiO_3$-Terfenol-D.[10-12] All these composite materials have been exclusively studied in bulk form.

Interestingly, the materials made in the form of superlattice structure yielded unusual electrical and magnetic transport properties that cannot be obtained by classical solid-state chemistry route. Thus it is possible to construct superlattices whose structure consists of alternating ferroelectric and ferromagnetic layers, and one can investigate the coupling between the two properties. However these types of superlattice structures have rarely been investigated. The present work explores the possibility of fabricating one such superlattice structure. We have chosen the insulating $Pr_{0.85}Ca_{0.15}MnO_3$ (PCMO) as ferromagnetic layer and $Ba_{0.6}Sr_{0.4}TiO_3$ (BST) as ferroelectric layer. The PCMO is chosen to reduce the leakage current because its an insulator and BST is chosen to have minimum lattice mismatch with PCMO (~2 %). The complex impedance was measured, in the frequency range ($10-10^6$ Hz) at different magnetic field (0-7 T) in the range of temperature from 80 K to 300 K, on the superlattices of PCMO/BST. The superlattice exhibits ferromagnetic and ferroelectric transitions with 120 K and 140 K as the transition temperatures, respectively. The $PCMO_{10}/BST_9$ superlattice shows 40 % magnetoresistance (MR) in 7 T magnetic field at 80 K, which is an order of magnitude higher than the PCMO film of same thickness. The high MR could be explained by the magnetoelectric effect of the superlattice sample.

PCMO/BST superlattices were fabricated by pulsed-laser deposition using stoichiometric targets on $SrTiO_3$ (001) substrates at 720 °C in a flowing 100 mTorr oxygen atmosphere. Superlattices with individual BST layer thickness of 2 and 9 unit cells (u.c.) by keeping PCMO layer thickness as 10 u.c. were fabricated along with PCMO, whose thickness is the same as that in superlattices. The superlattice is composed of 25 repeated units of PCMO/BST bilayers with PCMO as the bottom layer and BST as the top layer. The samples were characterized by x-ray diffraction (XRD) using Seifert 3000P diffractometer (Cu K$\alpha$, $\lambda$ = 1.5406 Å). Magnetization (M) was measured as a function of temperature (T) and magnetic field (H) using a superconducting quantum interference device magnetometer (SQUID). The complex impedance was measured by a lock-in amplifier (Standford Research 850) in the frequency range ($1-10^6$ Hz). The sample was hold in a Physical Property Measurement System (PPMS) (from Quantum Design) which provides the temperature regulation (2-400K) and the magnetic field (up to 7 T). The co-axial wires were directly attached on a small sample holder with indium solder. The voltage was measured on two different contacts as in usual four terminal resistivity method in order to eliminate the contact impedance. A 1 volt AC-voltage ($V_0$) was applied in the ($1-1\times10^6$) Hz frequency range. The current was measured on an additional resistor $R_0$ (typically 10 ohms). The contributions of the measured impedance, the air permittivity, the contacts and the cables (resistive and inductive effects) were carefully removed in order to obtain the actual permittivity of the sample. We have used current-perpendicula-to-the-plane (CPP) geometry in all our measurements using $LaNiO_3$ as an electrode.

In Fig. 1a, we show the $\theta$-$2\theta$ XRD scan around the (002) fundamental peak (40° - 52° in $2\theta$) of $PCMO_{10}/BST_9$ superlattice. The denoted number *i* indicates the *ith* satellite peak. The presence of higher order satellite peaks adjacent to the main peak, arising from chemical modulation of multilayer structure, indicates that the films were indeed coherent heterostructurally grown. We have carried out the XRD simulation of the superlattice structure using *DIFFaX* program[13] and it is found that the experimentally measured peaks are in good agreement with the simulated one (see Fig. 1a). The full-width-at-half-maxima (*FWHM*) of the rocking curve, recorded around the fundamental (002) diffraction peak of the superlattice sample which is shown as an inset in Fig. 1a, is very close to the instrumental broadening (0.143°), indicating a good crystalline quality.

In order to characterize the ferromagnetic properties of the superlattice, we show magnetization hysteresis loop measured at 10 K for the two superlattice and PCMO samples in Fig. 1b. The inset in the figure shows the corresponding magnetization with temperature plot. We magnetize our samples in the in-plane direction which is found to be predominantly the easy axis of magnetization (not shown). From Fig. 1b, it is inferred that all the samples are ferromagnetic with the Curie temperature $T_C$ around 120 K.

In Fig. 2a, we show the real $V_R$ and imaginary $V_I$ parts of the voltage measured across the resistor $R_0$ (10 Ω) at 80 K, using a lock in amplifier. The schematic diagram of the measurement set-up is shown as an inset in Fig. 2a. The modulus of the complex impedance, |Z|, of the sample is derived from the measured voltages $V_{meas} = (R_0V_0)/(R_0+Z)$, where $R_0$=10

$\Omega$, and $V_0$ is the applied 1 volt AC voltage. Since $R_0 << Z$, the modulus of the complex impedance can be written:

$$|Z| = (R_0 V_0)/ |V_{meas}| = 10/\sqrt{(V_R^2 + V_I^2)}.$$

Fig. 2b shows the calculated |Z| as a function of frequency with a range of temperature near the $T_C$ (80 – 150 K). In all the measurements, the ac current intensity was chosen small enough to prevent any nonlinear or heating effects often observed in dc measurements. We have modeled an equivalent electronic circuit for the observed ac response, shown as an inset in Fig. 2b. Here, the overall sample Z can be represented by 25 units (the number of bilayers) of the circuit connected in series, where in each unit the resistance $R_1$ of the PCMO layer is in series with the parallel combination of the resistance $R_2$ and capacitance C of the ferroelectric BST layer. The complex impedance inferred from the equivalent circuit $Z = R_1 + 1/[(1/R_2) + j\omega C]$ yielded,

$$|Z|^2 =[(R_1 + R_2)^2 + (CR_1R_2\omega)^2]/[1 + (R_2C\omega)^2]$$

From the above expression, one can note that at low frequency, the limit is a constant value of $(R_1 + R_2)$ and at high-frequency, limit is the resistance $R_1$. Although the model is not giving perfect fit to the experimental data, because of the poor information about the interface, it is reasonably good near the two limiting frequency range. As seen from the Fig. 2b, below $T_C$ there exists a crossover with an order of magnitude decrease in Z. The crossover frequency is increasing with the temperature up to $T_C$ and above it. We did not see any crossover within our temperature range of measurement. At present, the significance of the crossover is not clearly understood.

The dielectric constant $e_r$ of the $PCMO_{10}/BST_9$ sample, derived from the real and imaginary parts of the measured complex impedance, with temperature is plotted in Fig. 3 along with the resistivity of the sample. A fairly good match between the measured dc resistivity (solid line) and the resistivity calculated from $R_1+R_2$ is observed. It confirms that our model is reasonable good at lower frequency limit. [As a representative example, in the inset of the Fig. 3, we show the real and imaginary parts of the Z along with the frequency limits used to derive the resistive and capacitance parts of the sample at 120 K.] The Fig. 3 clearly shows a peak in dielectric constant at 140 K, where BST undergoes a ferroelectric to paraelectric transition. The high value of the dielectric constant (~100), often a good indicator of the presence of ferroelectricity, and the ferroelectric to paraelectric transition strongly suggests that the superlattice could have the ferroelectric properties.

In order to see the magnetoelectric effect on our samples, we measured complex impedance of the superlattices by applying external magnetic field up to 7 T at three different temperature of interest, i.e., at 80, 130, and 300 K. The resultant modulus of impedance for the samples $PCMO_{10}/BST_9$, $PCMO_{10}/BST_2$, and PCMO are shown in Fig. 4a, 4b, and 4c, respectively. The following points are interesting to note. First, the crossover exists only for the superlattice samples and not for the PCMO film. Second the crossover is observed at lower frequency with higher BST layer thickness and is shifted to higher frequency with decrease in BST layer thickness in the superlattice samples. Third, the change in impedance with magnetic field is high for higher BST layer thickness compared to the superlattice with lesser BST layer thickness, and the PCMO film. We have calculated the magnetoresistance (MR), i.e., the change in impedance with magnetic field, at two different frequency limit (at $\omega \rightarrow 0$ and $\omega \rightarrow \propto$) where the contributions are only from the resistive parts. For the same thickness of the film the sample $PCMO_{10}/BST_9$ shows a higher value of negative MR, 40 %, than the PCMO film (4 %) at lower frequency, where the contributions mostly come from the combination of the resistance $R_1+R_2$. At high frequency limit, all the samples show nearly 3 to 4 % of MR where the contribution comes from the resistive part $R_1$ of the PCMO layer.

Since the Fig. 1b and Fig. 3 suggest the coexistence of ferromagnetic and ferroelectric properties in $PCMO_{10}/BST_9$, the change in |Z| with applied magnetic field could be explained

by the associated magnetoelectric effect in the superlattice. The existence of negative differential resistance with an electric field in a similar compound has already been reported in the literature.[14] Thus the large MR can be attributed to the negative differential resistance due to the charges at the interface from the piezoelectric BST layer induced by the magnetostriction of the ferromagnetic PCMO layer at high applied magnetic field (7 T).[15]

In conclusion, mutiferroic superlattices composed of ferromagnetic $Pr_{0.85}Ca_{0.15}MnO_3$ and ferroelectric $Ba_{0.6}Sr_{0.4}TiO_3$ were made by pulsed laser deposition. The resistive and capacitance parts of the superlattices were derived from the complex impedance measured at different temperatures with applied magnetic field using a special experimental set-up. The magnetization and the dielectric constant measured as a function of temperature indicates that the superlattice could have the coexistence of both ferromagnetic and ferroelectric properties. The magnetoresistance of 40 % was observed at 80 K for the superlattice with high BST layer thickness. The observed high magnetoresistance could be attributed to the associated magnetoelectric effect.

Acknowledgements:

We thank J. Lecourt for the target preparation and F. Ferreira for preparation of special probe which we used in the PPMS. One of the author (P.M) acknowledges the Ministere de la Jeunesse et de l'Education Nationale for his fellowship (2003/87).

**Reference**

[1]H. Schmid, Ferroelectrics **62,** 317 (1994).

[2]H. Schmid; Ferroelectrics **221**, 9 (1999).

[4]J. Wang, J. B. Neaton, H. Zheng, V. Nagarajan, S. B. Ogale, B. Liu, D. Viehland, V. Vaithyanathan, D. G. Schlom, U. V. Waghmare, N. A. Spaldin, K. M. Rade, M. Wuttig, and R. Ramesh, Science **299**, 1719 (2003).

[4] N. A. Hill, J. Phys. Chem. B **104**, 6694 (2000).

[5]J. van Suchtelen, Philips Res. Rep. **27**, 28 (1972).

[6]J. Van Den Boomgaard, D. R. Terrell, and R. A. J. Born, J. Mater. Sci. **9**, 1705 (1974).

[7] J. Van Den Boomgaard, A. M. J. G. van Run, and J. Van Suchtelen, Ferroelectrics **14**, 727 (1976).

[8] J. Van Den Boomgaard, A. M. J. G. van Run, and J. van Suchtelen, Ferroelectrics **10**, 295 (1976).

[9]G. Srinivasan, E. T. Rasmussen, J. Gallegos, R. Srinivasan, Y. I. Bokhan, and V. M. Laletin, Phys. Rev. B **64**, 214408-1 (2001).

[10]J. Ryu, A. V. Carazo, K. Uchino, and H.-E. Kim, Jpn. J. Appl. Phys. **40**, 4948 (2001).

[11]K. Mori and M. Wuttig, Appl. Phys. Lett. **81**, 100 (2002).

[12]S. Dong, J.-F. Li, and D. Viehland, Appl. Phys. Lett. **83**, 2265 (2003).

[13]See at http://ccp14.sims.nrc.ca/ccp/ccp14/ftp-mirror/diffax/pub/treacy/DFFaX_v1807.

[14]S. Mercone, A. Wahl, Ch. Simon, and C. Martin, Phys. Rev. B **65**, 214428 (2002).

[15]G. Srinivasan, E. T. Tasmussen, B. J. Levin, and R. Hayes, Phys. Rev. B **65**, 134402 (2002).

**Figure Captions:**

FIG. 1. (a) Observed and calculated $\theta - 2\theta$ XRD scan recorded around the (002) reflection of $SrTiO_3$ substrate for $(PCMO_{10}/BST_9)_{25}$ superlattice. The symbol *i* indicate the number of satellite peak. The inset shows the rocking curve measured around the fundamental diffraction peak of the superlattice. (b) Magnetic hysteresis loop measured at 10 K for $PCMO_{10}/BST_9$, $PCMO_{10}/BST_2$, and PCMO samples. The inset shows the corresponding temperature dependent magnetization curve.

FIG. 2. (a) Real and imaginary parts of the measured voltage across the 10 Ω resistor for $PCMO_{10}/BST_9$. The inset show the schematic diagram of the experimental set-up. (b) The modulus of complex impedance with frequency calculated at different temperature. The inset shows the equivalent circuit diagram modeled for the observed ac response of $PCMO_{10}/BST_9$.

FIG. 3. The variation of dielectric constant and the resistivity with temperature for $PCMO_{10}/BST_9$ sample. The solid line indicates the measured dc resistivity and the open circle indicates resistivity calculated from $R_1+R_2$. The inset shows the real and imaginary parts of the complex impedance used to calculate the capacitance and $R_1+R_2$ of the sample at 120 K.

FIG. 4. (a) Modulus of complex impedance versus frequency with applied magnetic field at different temperature for (a) $PCMO_{10}/BST_9$, (b) $PCMO_{10}/BST_2$, and (c) PCMO samples.

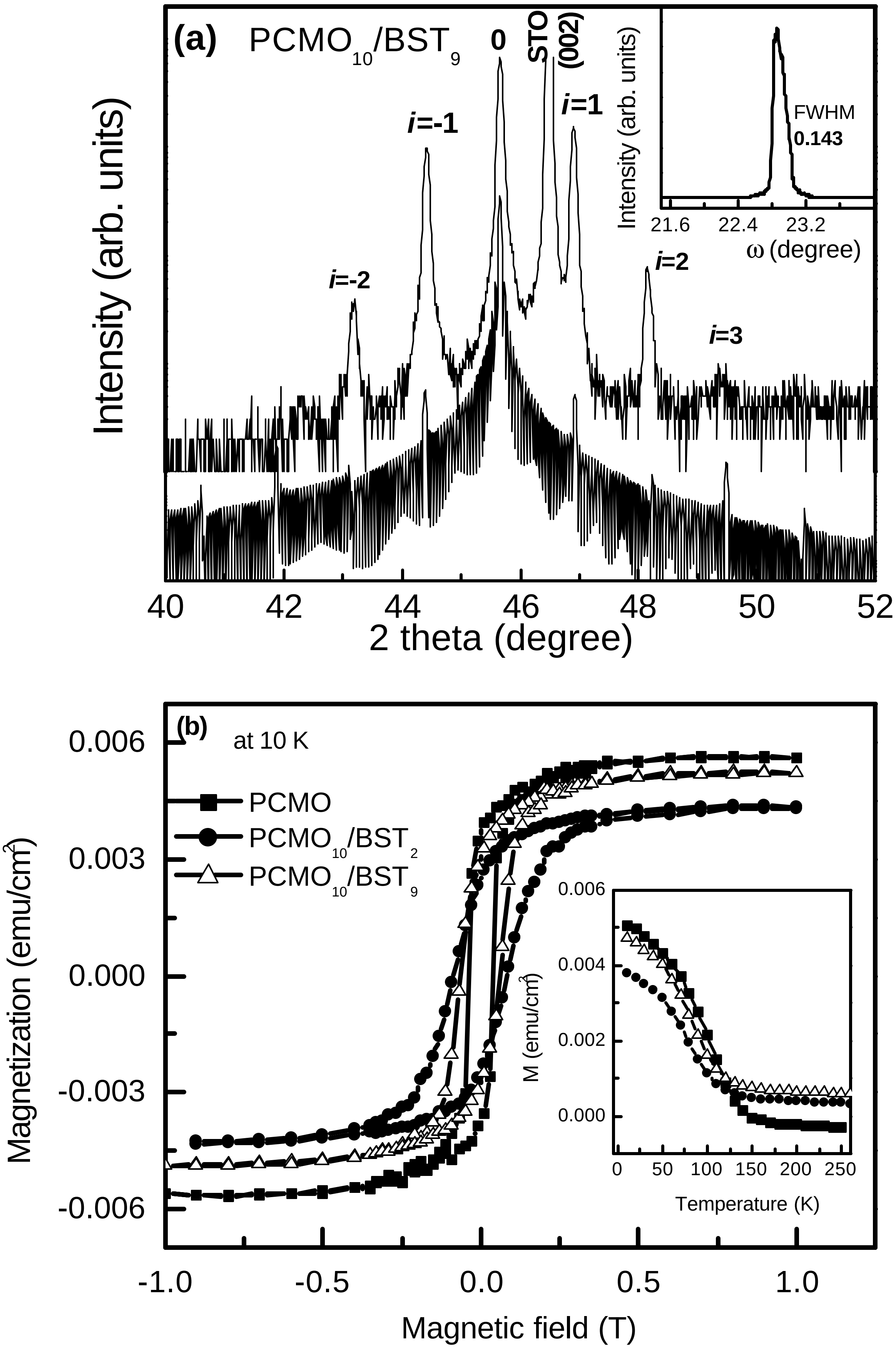

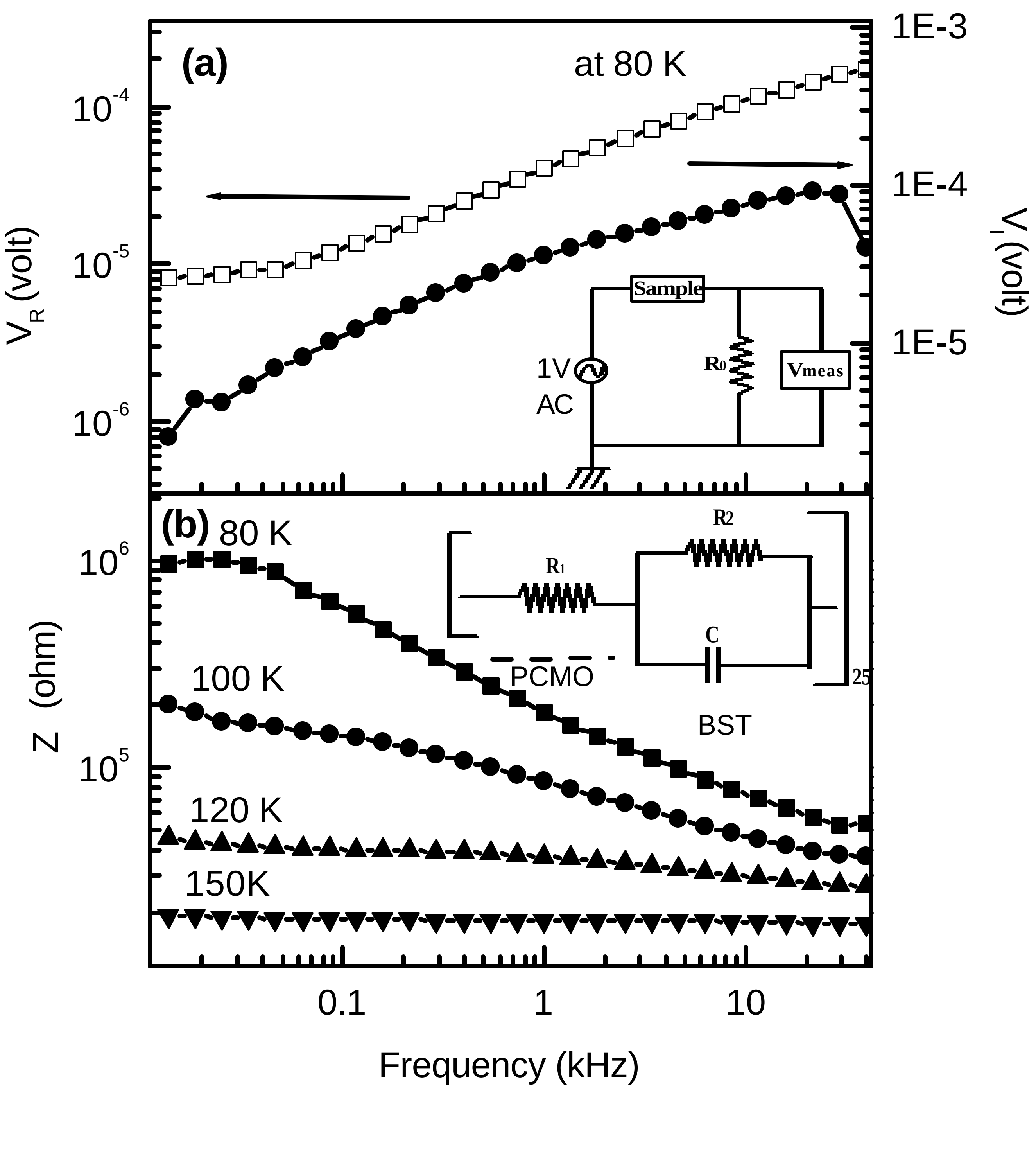

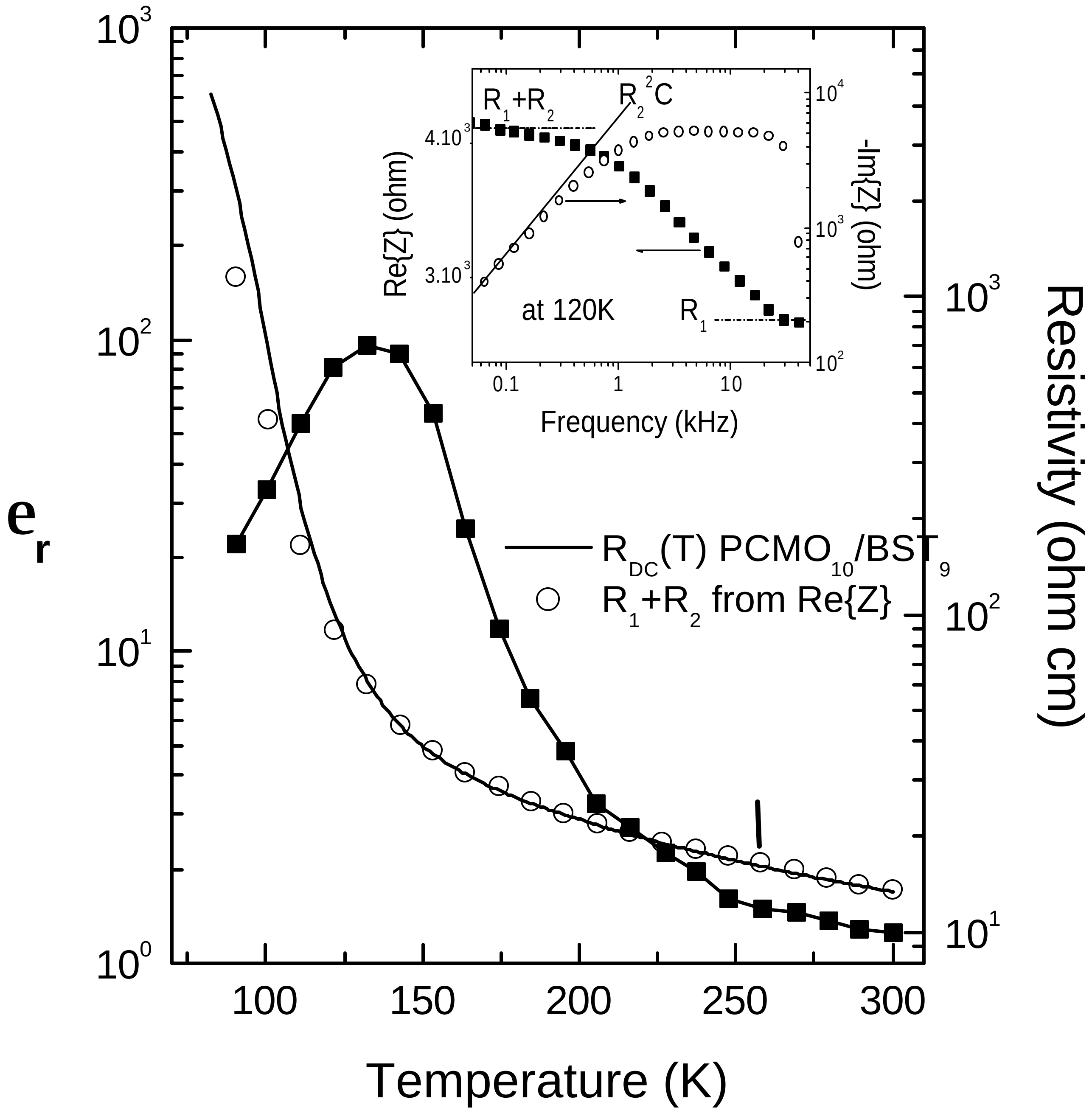

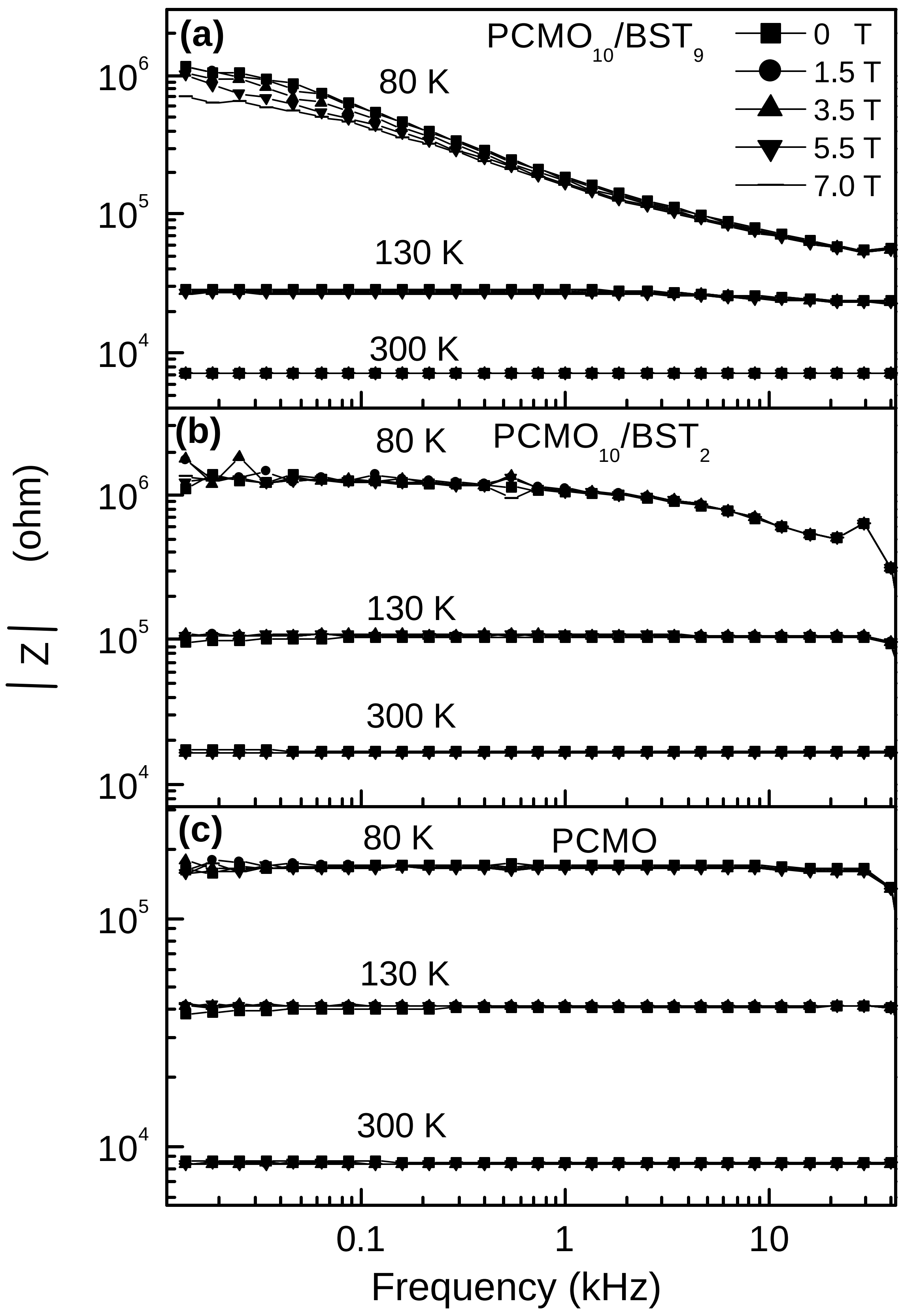